\input{aipcheck}

\documentclass[
    ,final            
  ]
  {aipproc}
\usepackage[usenames,dvipsnames]{color} 
\usepackage{multirow}
\usepackage{amsmath}
\usepackage{mciteplus}

\layoutstyle{8x11single}

\newcommand{\ii}{i}                    
\newcommand{\de}{d}                    
\newcommand{\bm}{}

\begin{document}

\title{Transverse-momentum-dependent parton distributions (TMDs)}


\author{Alessandro Bacchetta}{
  address={Dip. di Fisica Nucleare e Teorica, Universit\`{a} di 
Pavia, and INFN, Sez. di Pavia, via Bassi 6, I-27100 Pavia, Italy}
}

\begin{abstract}
Transverse-momentum-dependent parton distributions (TMDs) provide three-dimensional
images of the partonic structure of the nucleon in momentum space. We made impressive
progress in understanding TMDs, both from the theoretical and experimental
point of view. This brief overview on TMDs is divided in two parts: in the
first, an essential list of achievements is presented. In the second, a
selection of open questions is discussed. 
\end{abstract}

\maketitle


Providing detailed maps of the nucleon partonic structure is one of 
the main goals of subnuclear physics. 
Such maps are the starting steps
toward the comprehension of confined QCD dynamics. They are essential for the
interpretation of any high-energy process involving hadrons.

After about forty years of study, we have reached a good knowledge of (some) 
collinear parton
distribution functions (PDFs). They map the partonic structure in a
monodimensional momentum space, i.e., as a function
of partonic longitudinal momentum. ``Longitudinal'' is defined as 
parallel to the hard probe, which must be always present in hard processes
where PDFs are observed (e.g., in DIS the 
photon with virtuality $Q^2$).

TMDs (an acronym for Transverse Momentum Distributions or Transverse
Momentum Dependent parton distribution functions) are natural extensions of
standard PDFs: they are three-dimensional maps of the partonic structure in
momentum space. They depend not only on the longitudinal momentum, but also 
on the transverse momentum. 

\begin{figure}
\begin{tabular}{cc}
  \includegraphics*[height=4.7cm]{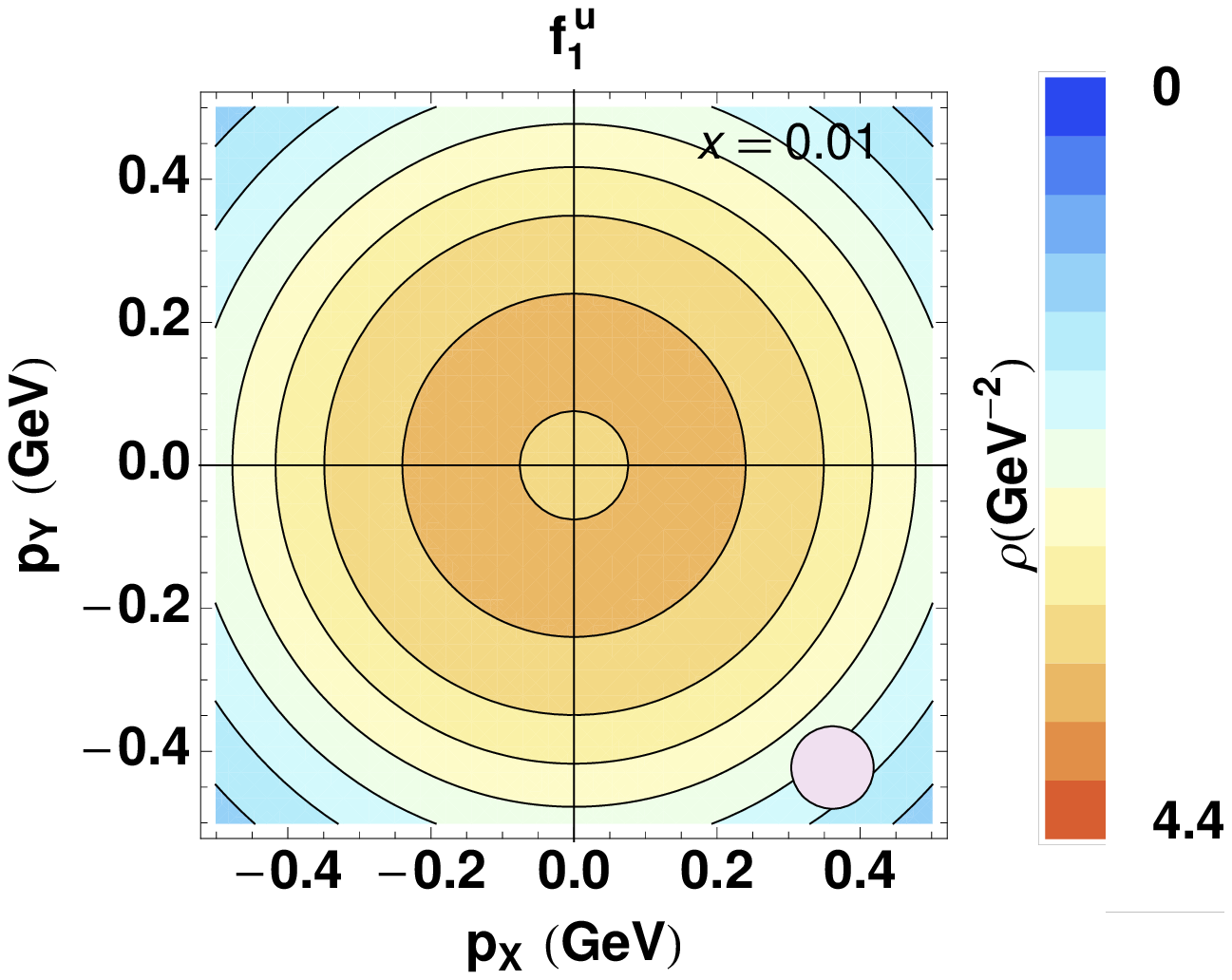}
&
  \includegraphics[height=4.7cm]{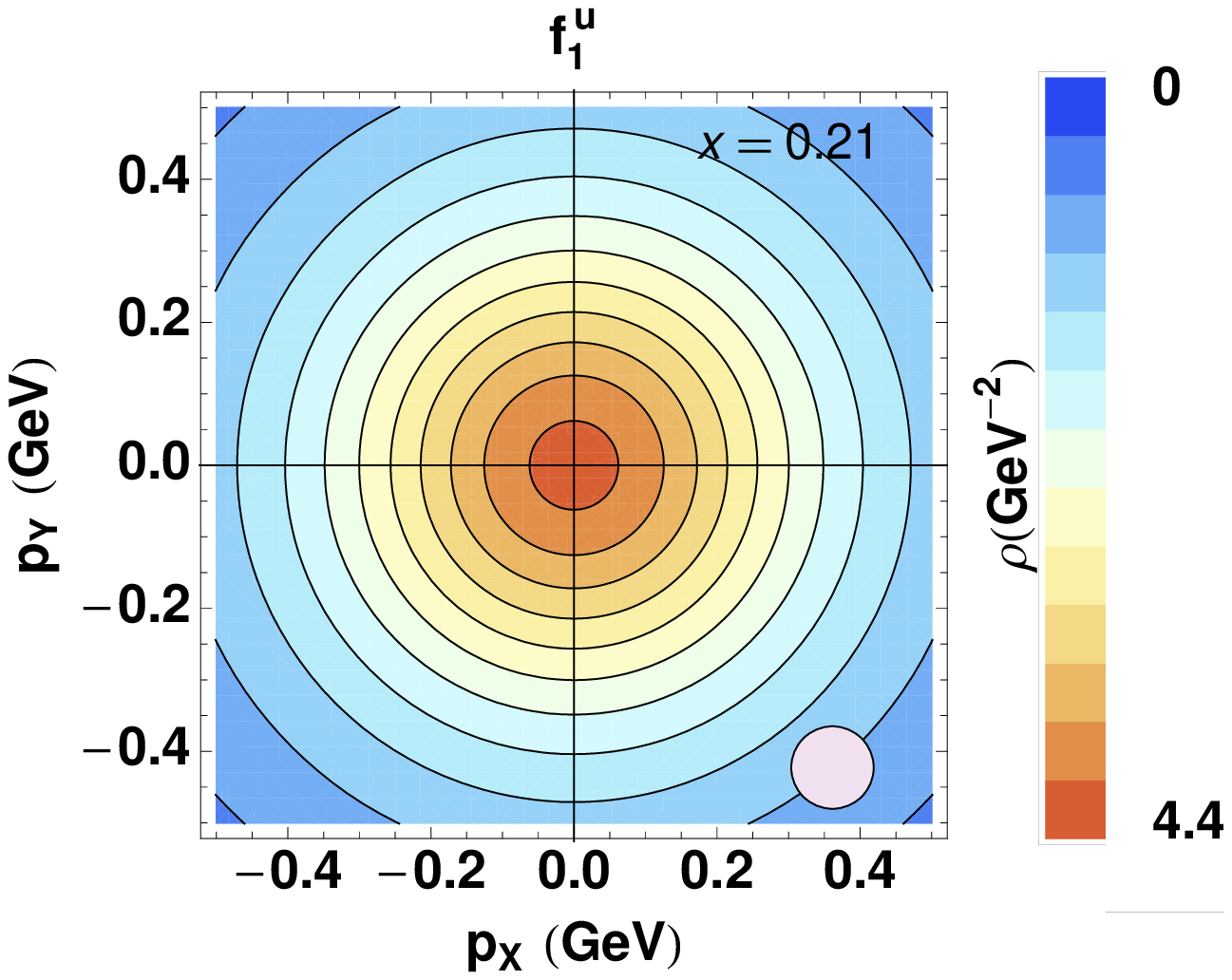}
\\
  \includegraphics*[height=4.7cm]{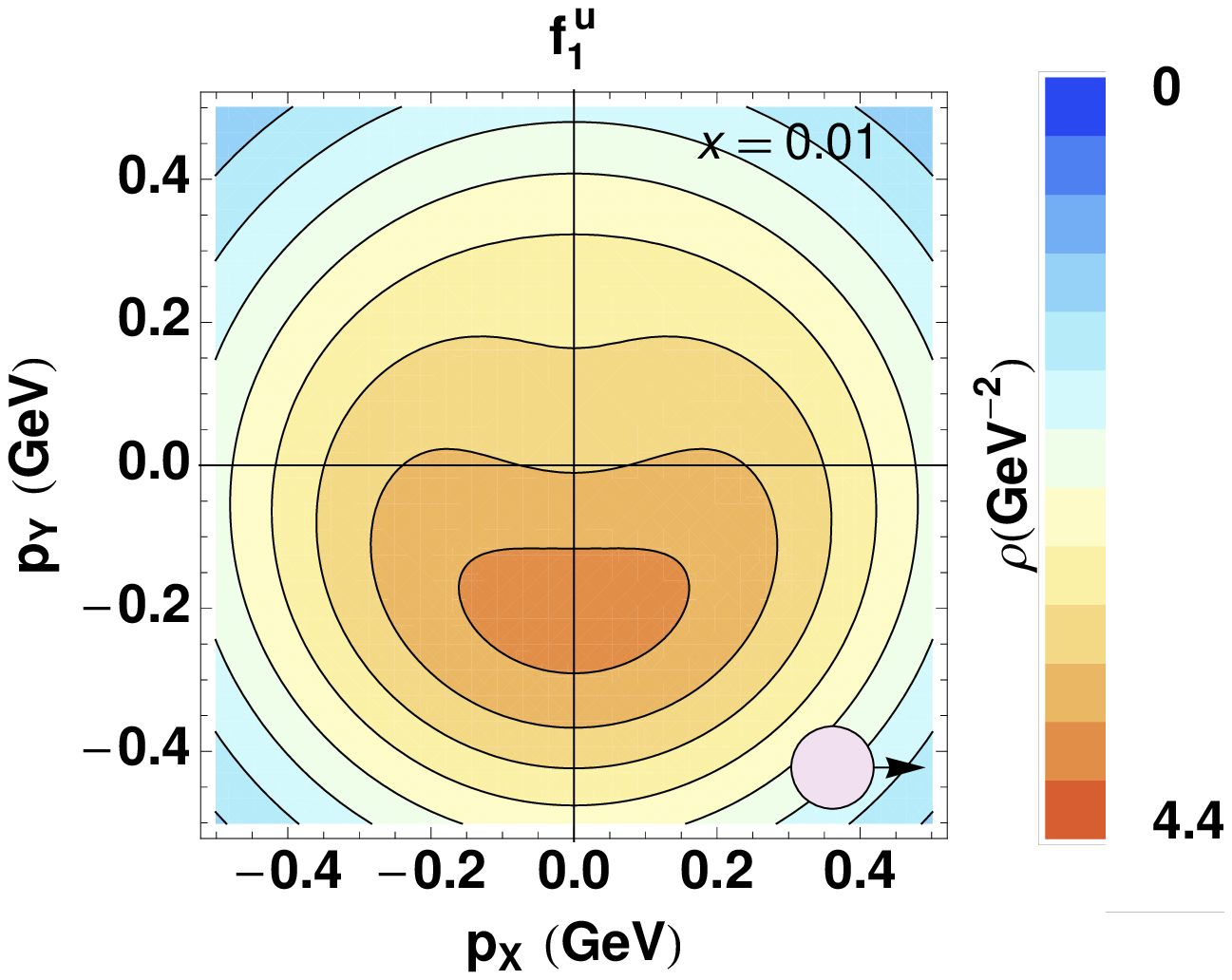}
&
  \includegraphics[height=4.7cm]{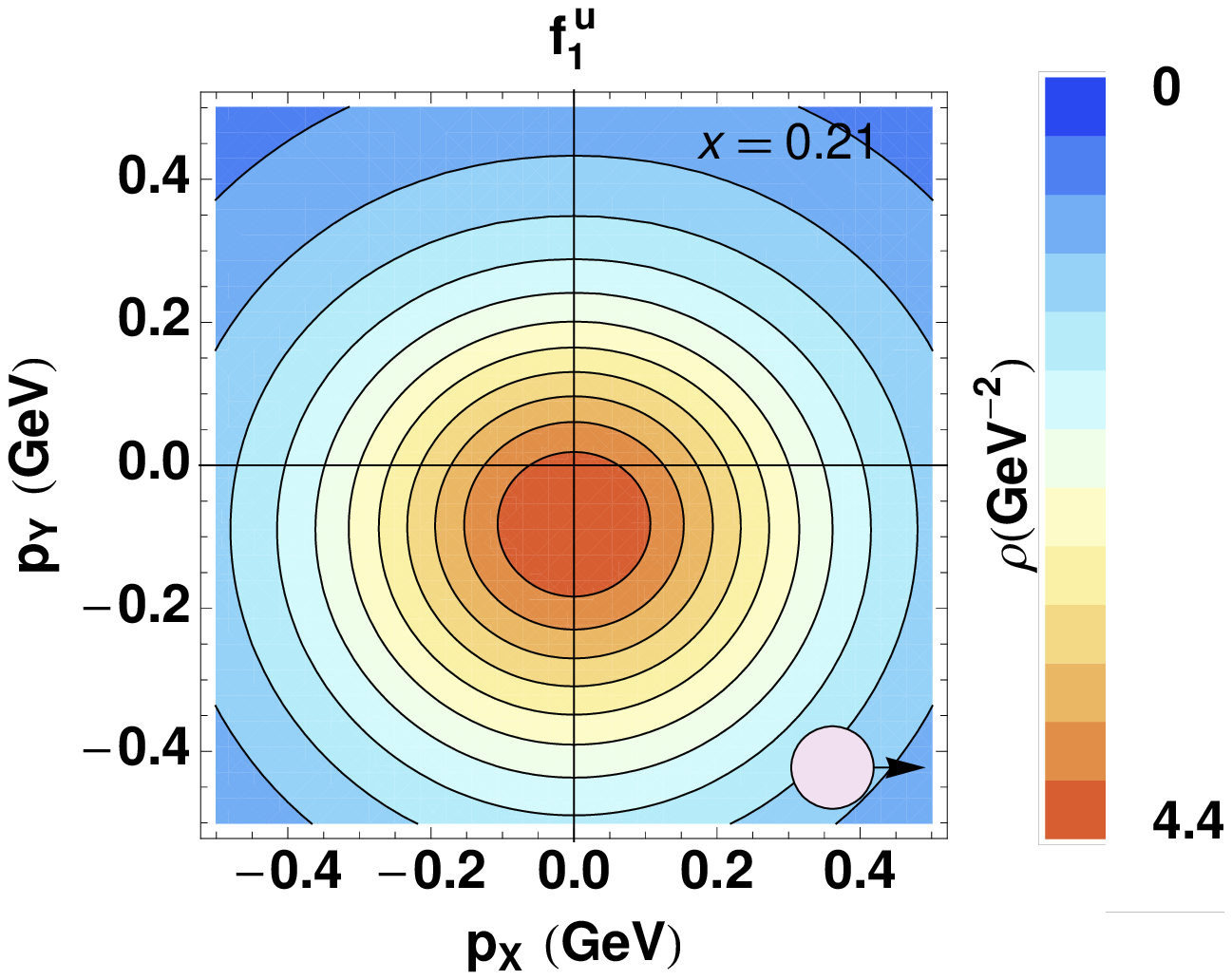}
\end{tabular}
  \caption{Momentum-space tomographic ``images'' of the up quarks in a nucleon
    obtained from a model calculation of TMDs~\cite{Bacchetta:2008af,*Bacchetta:2010si}. 
    The nucleon is assumed to move into the page. The circle with the arrow
    indicates the nucleon and its spin orientation. The distortion in the
    lower panels is due to the Sivers function.}
\label{f:tomo}
\end{figure}

Ultimately, the knowledge of TMDs will allow us to build tomographic images
of the inner structure of the nucleon in momentum space, complementary to the
impact-parameter space tomography that can be achieved by studying generalized
parton distribution functions (GPDs). An example of tomographic images of
the nucleon based on a model calculation of 
TMDs~\cite{Bacchetta:2008af,*Bacchetta:2010si} is
presented in Fig.~\ref{f:tomo}.  

\renewcommand{\arraystretch}{1}

The information contained in TMDs is as fundamental as that contained in standard
PDFs. They reveal crucial aspects of
the  
dynamics of confined partons, they can be extracted from experimental data,
and allow us to make prediction for
hard-scattering experiments involving nucleons.

Our understanding of TMDs and their extraction from data has made giant steps
in the last years, thanks to new theoretical ideas and experimental
measurements (see Ref.~\cite{Barone:2010zz} for a recent review). 
In the near future, more experimental data are expected from
HERMES, COMPASS, BELLE and JLab. A high-energy, high-luminosity, polarized
electron-proton collider, such as the envisaged EIC project, would undoubtedly be a
precision machine for the study of TMDs.

\section{What we know about TMDs}

Similarly to 
standard collinear PDFs, it is essential to define TMDs in a formally clear way,
through the proof of factorization theorems. TMDs appear when factorizing 
semi-inclusive processes. For instance, while totally inclusive DIS can be
described introducing collinear PDFs, TMDs appear in semi-inclusive DIS if
the transverse momentum of one outgoing hadron, $P_{h \perp}$, is
measured. 

Dealing with semi-inclusive processes pushes the difficulty of proving
factorization theorems to a higher level of complications. TMD factorization
is in fact a challenging arena where many of the simplifications used
in collinear factorization cannot be applied. Nevertheless, 
factorization for semi-inclusive DIS has been worked out explicitly at
leading twist (twist 2) and one-loop order~\cite{Collins:1981uk,*Ji:2004wu}.  
For instance, the structure function $F_{UU,T}$ in
the region $P_{h \perp}^2 \ll Q^2$ can be expressed as
\begin{align}
F_{UU,T} &=
\bigl| H\bigl(x \zeta^{1/2}, z^{-1}\zeta_{\smash{h}}^{1/2} , \mu_F\bigr)
\bigr|^2 \,
\sum_a x\, e_a^2
\int \de^2 \bm{p}_T\, \de^2 \bm{k}_T^{}\, \de^2 \bm{l}_T^{}\,
\nonumber \\
& \times
  \delta^{(2)}\bigl(\bm{p}_T - \bm{k}_T^{} +\bm{l}_T^{} - \bm{P}_{h \perp}^{}/z
  \bigr)\,
f_1^a(x,p_T^2;\zeta, \mu_F)\, D_1^a(z,k_T^2;\zeta_h,\mu_F)\, U(l_T^2;\mu_F) \,.
\label{FUUTconv}
\end{align}
The formula contains the (calculable) hard scattering factor $H$, the
transverse-momentum-dependent PDFs and fragmentation functions,
and 
the soft factor $U$, a nonperturbative,
process-independent object. However, it should be also possible to include the soft
factor into the PDFs and fragmentation functions by redefining
them in a suitable way~\cite{Collins:2000gd}.

A schematic definition of quark TMDs is 
(taking, e.g., the
unpolarized distribution of a quark with flavor $a$)
\begin{equation}  
f_1^a
(x,p_T^2; \zeta, \mu_F)= \int 
        \frac{\de \xi^- \de^2 \bm{\xi}_T}{(2\pi)^{3}}\; 
 e^{\ii p \cdot \xi}\,
       \langle P|\bar{\psi}^a(0)\,
{\cal L}^{v \dagger}_{(\pm\infty,0)}\,
\gamma^+
{\cal L}^{v}_{(\pm\infty,\xi)}\,
\psi^a(\xi)|P \rangle \bigg|_{\xi^+=0}.
\label{e:phi} 
 \end{equation} 
The Wilson lines, ${\cal L}$, guarantee the gauge invariance of the TMDs. They
depend on the gauge vector $v$ and 
contain also components at infinity running in the transverse
direction~\cite{Ji:2002aa}. Beyond tree level, 
there
are several subtleties related to the treatment of soft and rapidity
divergences. Various ways to deal with these divergences have been
proposed~\cite{Collins:1981uk,*Ji:2004wu,
Collins:2003fm,*Cherednikov:2009wk}. 
In general, they require the introduction of a rapidity cutoff 
$\zeta$, on which the TMDs depend.

A remarkable property of TMDs is that the detailed shape of the Wilson line is
process-dependent. This immediately leads to the conclusion that TMDs are not
universal. However, 
for transverse-momentum-dependent fragmentation functions, the shape of
    the Wilson line appears to have no influence on physical
    observables~\cite{Collins:2004nx,*Yuan:2008yv,*Gamberg:2008yt,*Meissner:2008yf}.
    In SIDIS and Drell--Yan, the difference between the Wilson line consists
  in a simple direction reversal and leads to calculable effects, namely a
  simple sign reversal of all T-odd TMDs~\cite{Collins:2002kn}.

In general and in simplified terms, 
the study of TMD factorization requires a
deeper understanding of what happens when a quark is hit inside a nucleon,
with a particular attention to the infinitely many gluons that 
surround the quark,  
beyond the simple case when its transverse momentum is integrated over.

At present, especially for azimuthally-dependent structure functions,
 phenomenological analyses are often carried out using the
tree-level approximated expression (which corresponds to using
parton-model formulas for collinear PDFs and neglecting scaling violations).
\begin{align}
F_{UU,T} &=
\sum_a x\, e_a^2
\int \de^2 \bm{p}_T\, \de^2 \bm{k}_T^{}\,
  \delta^{(2)}\bigl(\bm{p}_T - \bm{k}_T^{} - \bm{P}_{h \perp}^{}/z
  \bigr)\,
f_1^a(x,p_T^2)\, D_1^a(z,k_T^2)\,.
\label{FUUTconv_tree}
\end{align}
The transverse-momentum dependence of the
partonic functions is usually assumed to be a flavor-independent
Gaussian~\cite{D'Alesio:2004up,*Schweitzer:2010tt}. The
tree-level approximation and the Gaussian
assumption are known to be inadequate at $P_{h \perp}^2
\gg M^2$, but they could effectively describe the physics at $P_{h \perp}^2
\approx M^2$. Especially for low-energy experiments, this is where the bulk of
the data is.

There is an extensive literature where the analysis is carried out to a higher
level of complication, but only 
for the specific case of unpolarized observables integrated over the azimuthal
angle of the measured transverse momentum. The analysis is usually
performed in the space of the Fourier-conjugate to $P_{h\perp}$  
($b$-space) in the Collins--Soper--Sterman (CSS)
framework~\cite{Collins:1984kg}.  
The
region of $P_{h \perp}^2 \gg M^2$, or $b^2\ll 1/M^2$, 
can be calculated perturbatively, but when $P_{h \perp}^2 \approx
M^2$ a nonperturbative component has to be introduced and its parameters must
be fitted to experimental data. This component is usually assumed to be a
flavor-independent Gaussian~\cite{Landry:2002ix}.

At present, we can make the conservative statement that unpolarized quark
TMDs seems to be well described by flavor-independent Gaussians with 
$\sqrt{\smash[b]{\langle p_T^2 \rangle}} \approx 0.4 - 0.8$ GeV, depending on the
kinematics. 

\begin{figure}
\begin{tabular}{ccc}
\includegraphics[height=3.5cm]{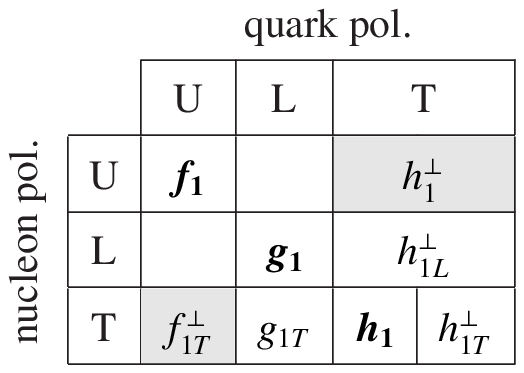}
&
\qquad \qquad
&
  \includegraphics[height=4.8cm]{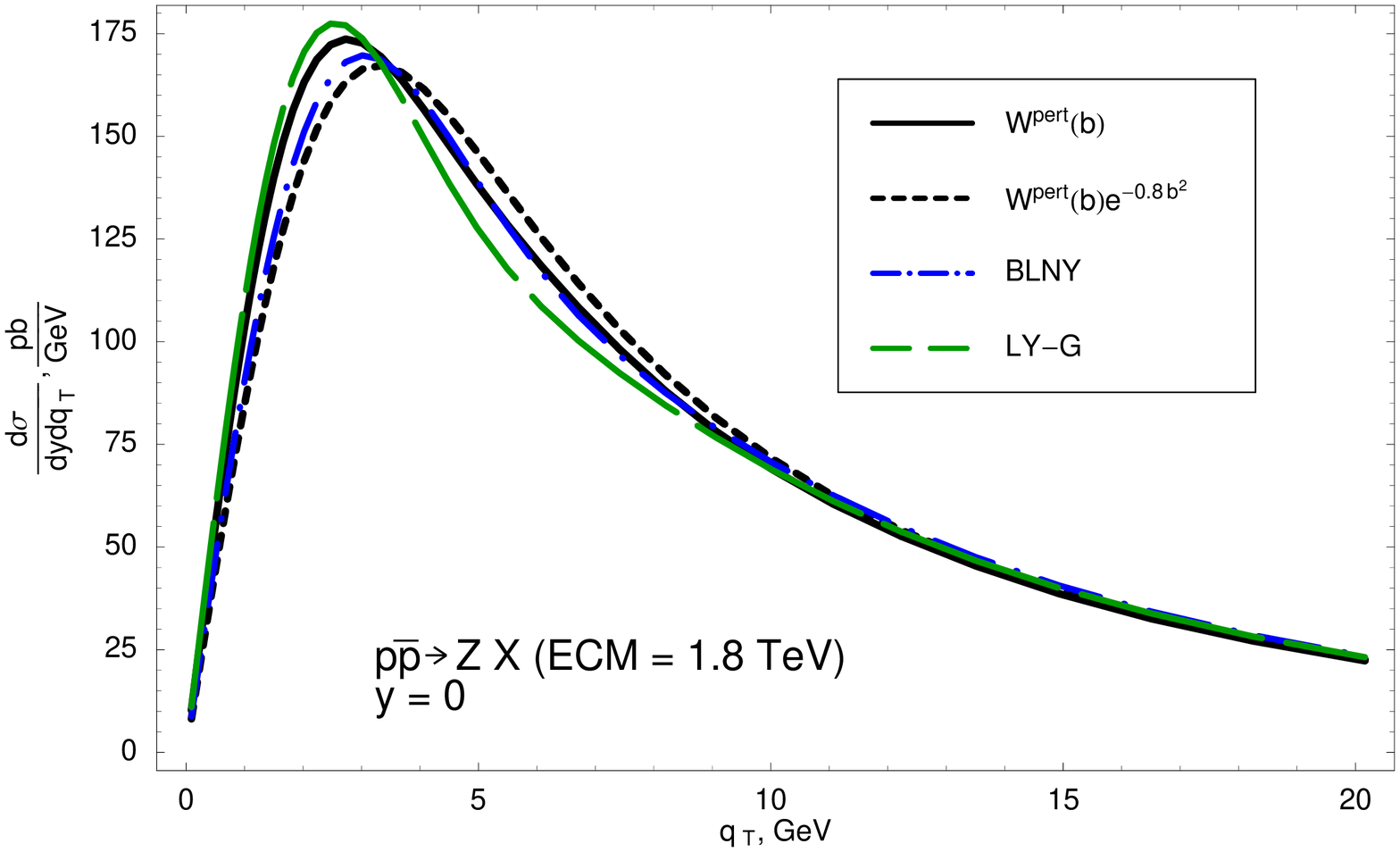}
\\
(a) && (b)
\end{tabular}
  \caption{
(a) Twist-2 transverse-momentum-dependent distribution functions.
The U,L,T correspond to unpolarized, longitudinally polarized and transversely
polarized nucleons (rows) and quarks (columns). Functions in boldface survive
transverse momentum integration. Functions in gray cells are T-odd. 
(b) The cross sections of $Z$ boson production
at the Tevatron, computed in the CSS formalism. 
The difference between the curves shows the impact of choosing 
different nonperturbative 
components for the TMDs. See Ref.~\cite{Nadolsky:2004vt} for details.
}
\label{f:ZqT}
\end{figure}

The knowledge of the details of the unpolarized TMDs has an impact also on
high-energy physics. In Fig.~\ref{f:ZqT}b, the cross section for $Z$ boson production
at the Tevatron is plotted~\cite{Nadolsky:2004vt}. The difference between the curves
originates from different models and fits of the nonperturbative component of
the TMDs. Apart from the details, the plot shows that the knowledge of TMDs is
essential for precision studies at the Tevatron.

Even the determination of a fundamental parameter of the Standard Model, the
mass of the $W$ boson, is affected by the uncertainties of the knowledge of
unpolarized TMDs. In Ref.~\cite{Aaltonen:2007ps}, the CDF collaboration
discussed several ways to fit the $W$
mass. According to the analysis, TMDs uncertainties 
generate an error of 3.9 MeV on the $W$
mass determination 
(the total systematic error is about 34 MeV).

When the spin of the nucleon and that of the quark are taken into account,
eight twist-2 TMDs can be introduced. They are listed in 
Fig.~\ref{f:ZqT}a. 
As with collinear PDFs, extracting TMDs calls for global fits to
semi-inclusive DIS, Drell--Yan, and $e^+e^-$-annihilation data. 

From the theoretical point of view, we know positivity bounds for
TMDs~\cite{Bacchetta:1999kz}, 
behavior at high transverse momentum~\cite{Ji:2006ub,*Bacchetta:2008xw,*Zhou:2009jm}, 
behavior at high
$x$~\cite{Brodsky:2006hj}.

Apart from the $f_1$ (unpolarized function) and the $p_T$-integral of $g_1$
(helicity distribution), we have at the moment some 
extractions of $h_1$ (transversity distribution),
$f_{1T}^{\perp}$ (Sivers function)~\cite{Anselmino:2008sga,Arnold:2008ap} and
$h_1^{\perp}$ (Boer--Mulders function)~\cite{Lu:2009ip,*Barone:2009hw}.

\begin{figure}
\begin{tabular}{ccc}
  \includegraphics[height=6.2cm]{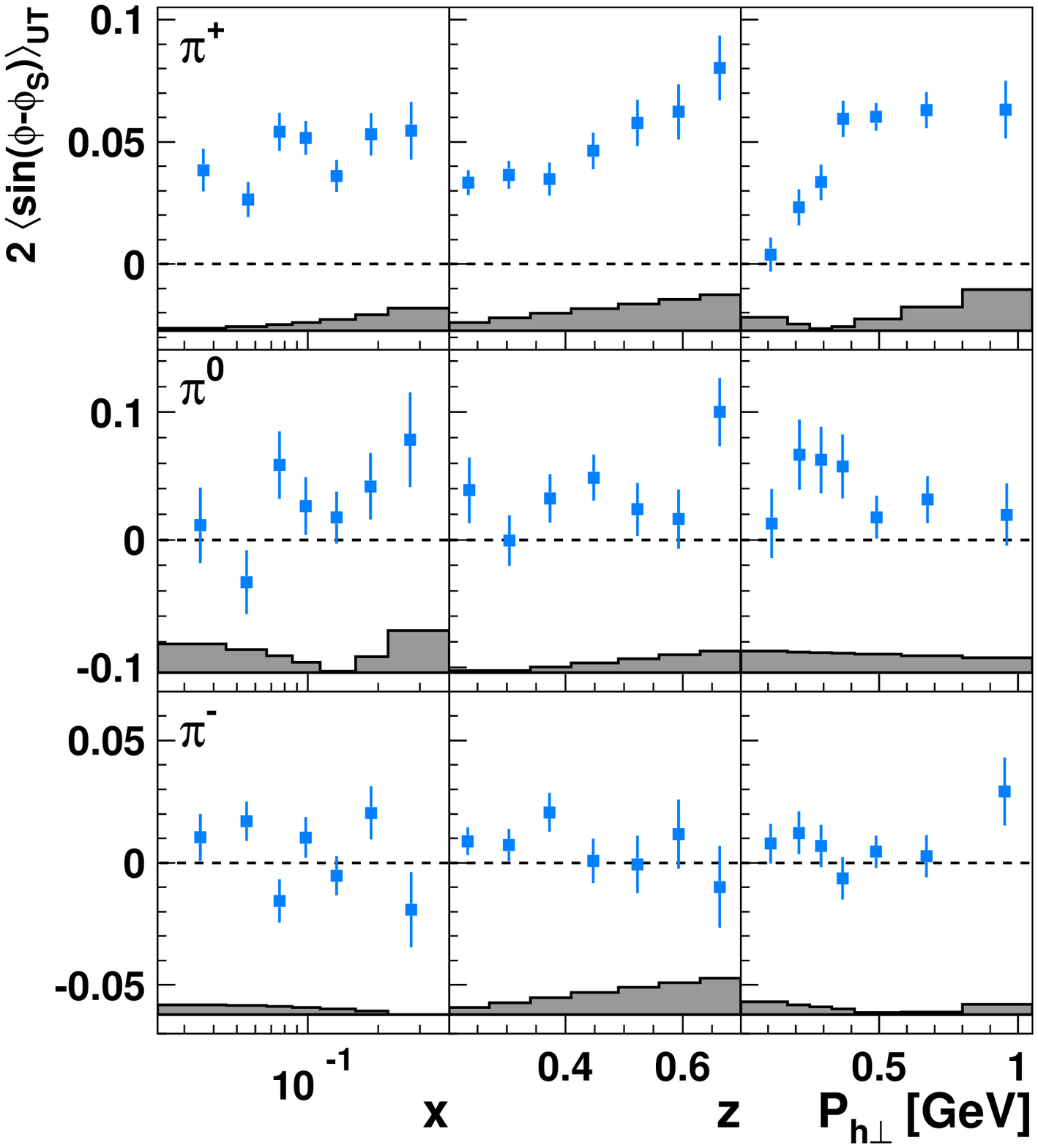}
&
\qquad
&
  \includegraphics[height=6.2cm]{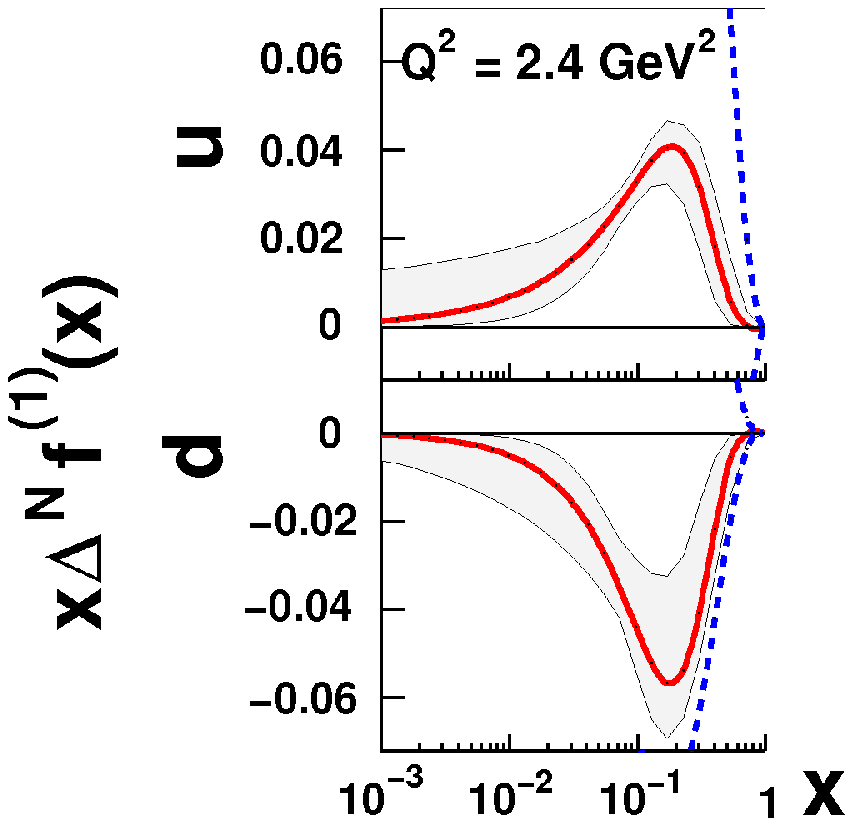}
\\
(a) && (b)
\end{tabular}
\caption{(a) Example of experimental data on the Sivers
  asymmetries~\cite{Airapetian:2009ti}. 
(b) Sivers distribution for up and down quark from a recent
extraction~\cite{Anselmino:2008sga}. 
} 
\label{f:siversfit}
\end{figure}

As illustrated in Fig.~\ref{f:tomo}, a nonzero Sivers function means that the
distribution of quarks in transverse momentum is affected by the direction of
the nucleon's spin. Thanks to the experimental measurements of the
HERMES~\cite{Airapetian:2009ti}  and
COMPASS~\cite{Alekseev:2008dn} 
collaborations, it has become possible to extract for the
first time the Sivers distribution function. Fig.~\ref{f:siversfit} shows a 
representative subset of the data and one
of the most recent parametrizations~\cite{Anselmino:2008sga,Arnold:2008ap}). 

The conclusions we can draw from these first studies are already
interesting: the Sivers function is nonzero, at least for up and down quarks;
the Sivers function for up quark is negative, while for down quark is positive
and could be even larger than that of up quarks. These findings are in
agreement with the idea suggested in Ref.~\cite{Burkardt:2002ks} of a 
 link between the anomalous magnetic moment, \(\kappa\), and the Sivers
 function. Assuming isospin symmetry and no strange quark contribution, we know that
 $\kappa_u \approx 1.673$ and $\kappa_d \approx -2.033$. Since quarks are
 subject to attractive final-state color interactions, this indeed translates in a
 negative up and a large, positive down Sivers, in qualitative agreement with
 some models~\cite{Courtoy:2008vi,*Courtoy:2008dn,*Pasquini:2010af}.

\section{What we still don't know about TMDs}

There are many essential questions that have not been addressed
sufficiently in the past. 

Concerning the formalism, there are a few alternative definitions of TMDs,
based on different ways to handle soft and rapidity divergences. It is at
present not clear if the different definitions are all formally acceptable, if
they can be seen as different ``schemes'' to define TMDs, and if some of them
are more convenient than others.

It is extremely important to clarify if and how TMDs can be studied in hadron-hadron
collisions to hadrons. The treatment of  Wilson lines in this case 
is more intricate and
recent studies lead to the conclusion that TMD factorization does not
work~\cite{Rogers:2010dm}. Efforts should be made to look for experimental
evidence of factorization-breaking effects, and to find ways to avoid
problems~\cite{}.

\begin{figure}
\begin{tabular}{ccc}
  \includegraphics[height=4.5cm]{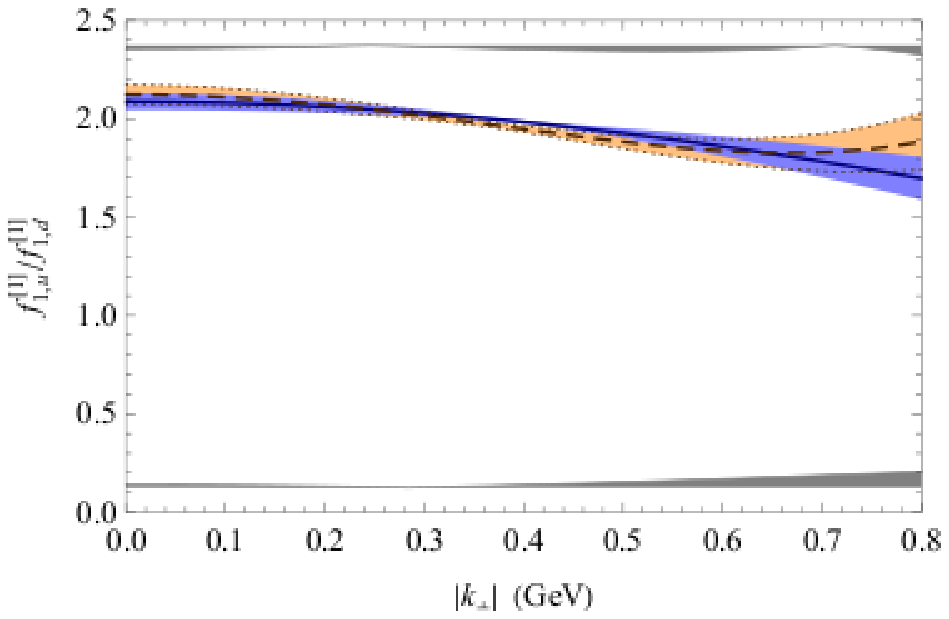}
&
\qquad
&
  \includegraphics[height=4.5cm]{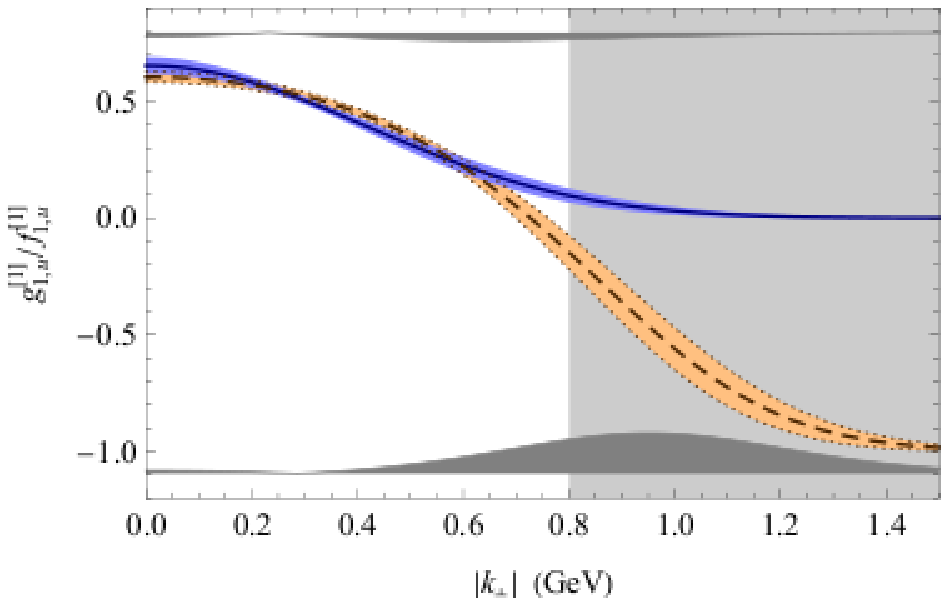}
\\
(a) && (b)
\end{tabular}
\caption{First hints on TMD behavior based on lattice QCD
  computations~\cite{Musch:2010ka}.  (a)
  Ratio between the TMD of up and down
  quarks from two different fits to lattice QCD data: the non-constant ratio 
  suggests that up and down
  quark have different transverse momentum distributions. (b) Ratio between
  the helicity distribution and the unpolarized distribution for up quarks:
  the non-constant ratio suggests that quarks with different spin orientation
  have different transverse momentum distributions. 
}
\label{f:lattice}
\end{figure}

For what concerns our phenomenological knowledge of TMDs,
we do not know whether the transverse momentum distribution is different for
different quark flavors and for gluons.
We know that collinear PDFs are flavor dependent: not only their
normalization is different, but also their shape. Their transverse spatial
distributions, as reconstructed from form factor measurements, are also sharply
different (see, e.g., \cite{Miller:2007uy}). Some models predict different
shapes~\cite{Bacchetta:2008af,*Bacchetta:2010si,Wakamatsu:2009fn}. 
There is no fundamental reason to 
believe that their transverse momentum distributions should be equal. 
However, up to now they have been assumed to be exactly the same in all
phenomenological analyses. There are in fact already some indications that
this might not be a good assumption, both from
experiments~\cite{Mkrtchyan:2007sr}
 and lattice-QCD calculations~\cite{Musch:2010ka} (see Fig.~\ref{f:lattice}a). 

We don't know enough about the detailed shape of TMDs. In all studies, it is 
assumed to be Gaussian. However, this is a choice based on convenience and
tradition, but not on a fundamental reason. There is no model
calculation of TMDs that predicts a pure Gaussian shape. 
The
presence of nonzero orbital angular momentum actually implies that the
shape cannot be a simple Gaussian. 
In nonrelativistic quantum mechanics, it is well known that
wave-functions with orbital angular momentum vanish at zero momentum. 
This feature is reflected also in TMDs:
contributions from partons with nonzero angular momentum have to vanish at
zero transverse momentum and therefore 
cannot be described by a simple Gaussian.
We can assume that the wave-function of the
proton contains quarks with $L_z=0$ ($s$ wave) and $L_z=1$ ($p$ wave), where
$L_z$ is defined in the Jaffe--Manohar way. Then we know that
\begin{equation}
\lim_{p_T \to 0}
\vert \psi_{p-\rm{wave}}\vert^2 
\propto
p_T^2 / M^4  
\end{equation}  
The full TMD will be the sum of the squares of different components of the
wave-function, making it difficult to identify the contributions with nonzero
$L$.  
In any case, 
a downturn of a TMD at small transverse momentum may signal the
presence of nonzero orbital angular momentum. 
While this effects could barely
be visible in unpolarized TMDs, certain combinations of polarized TMDs could
filter out more clearly the configurations with nonzero orbital angular
momentum.

%

Apart from the details of their shape, all the TMDs that are not boldface
in Fig.~\ref{f:ZqT}a vanish in the absence of
orbital angular momentum due to angular momentum conservation. Measuring any
one of them to be nonzero is already a indisputable indication of the
presence of partonic orbital angular momentum. We know already from other
sources (e.g., the measurement of nucleons' anomalous magnetic
moments) that partonic orbital angular momentum is not zero, however TMDs have
the advantage that they can be flavor-separated and that they are $x$
dependent.  Thus, they allow us to say if
orbital angular momentum is present for each quark flavor and for gluons, and 
at each
value of $x$.

If stating that a fraction of partons have nonzero orbital
angular momentum is relatively simple, 
it is not easy to make a quantitative estimate
of the net partonic orbital angular momentum using TMDs. 
Any statement in this direction is bound to be model-dependent.
Generally speaking, TMDs have to be computed in a model and the parameters of
the 
model have to be fixed to reproduce the TMDs extracted from data.  
Then, the
total orbital angular momentum can be computed in the model. 
Unfortunately, it is 
possible that two models describe the data equally well, but
give two different values for the total orbital angular momentum.

\begin{figure}
\begin{tabular}{ccc}
  \includegraphics[height=5cm]{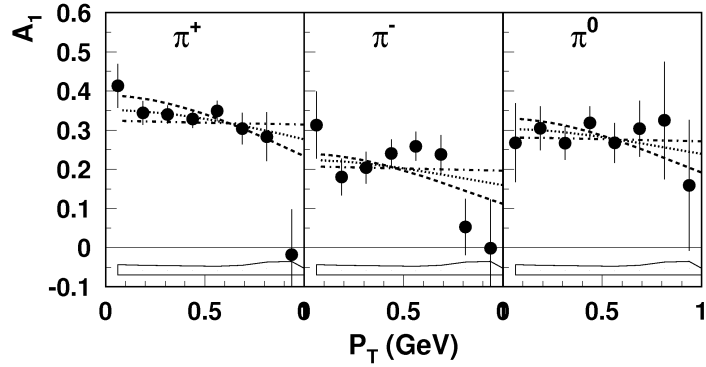}
\end{tabular}
\caption{Data from the CLAS collaboration showing the transverse-momentum
  dependence of the double longitudinal spin asymmetry
  $A_1$~\cite{Avakian:2010ae}. The dependence would be flat 
  if TMDs for quarks with opposite helicity were
  the same. 
} 
\label{f:ALL_pt}
\end{figure}

Another aspect that requires further investigations is to which extent 
transverse momentum distribution is influenced by the spin of the quarks and
of the nucleon. For instance, it may be possible that the average transverse
momentum of quarks with spin
antiparallel to the nucleon is larger than that of quarks with spin parallel
to the nucleon. Also in this case there is already some evidence from 
experiments~\cite{Avakian:2010ae} (see Fig.~\ref{f:ALL_pt})
 and lattice-QCD calculations~\cite{Musch:2010ka} (see
 Fig.~\ref{f:lattice}b). 

Finally, it is worth mentioning that during this conference the first
measurement connected to the so-called ``worm-gear'' TMD $g_{1T}$ has been
reported~\cite{Jiang:2010,*e06-010} 
and seems to confirm the sign of the distribution predicted by
lattice QCD studies~\cite{Hagler:2009mb}.

In summary, TMDs open new dimensions in the exploration of the partonic
structure of the nucleon. They require challenging extensions of the standard
formalism used for collinear parton distribution functions, leading 
 us to a deeper understanding of QCD. Among other things, 
they give evidence of the presence of
partonic orbital angular momentum and, with model assumptions, they can help
constraining its size.




\bibliographystyle{aipprocM}   

\bibliography{mybiblio}

\IfFileExists{\jobname.bbl}{}
 {\typeout{}
  \typeout{******************************************}
  \typeout{** Please run "bibtex \jobname" to optain}
  \typeout{** the bibliography and then re-run LaTeX}
  \typeout{** twice to fix the references!}
  \typeout{******************************************}
  \typeout{}
 }

\end{document}